\documentclass[journal]{IEEEtran}

\ifCLASSINFOpdf
\else
\fi

\usepackage{booktabs} 
\usepackage{caption} 

\usepackage{setspace, amsmath, amssymb, url, lscape, subfigure, algorithmic, multirow, pslatex, listings, verbatim, alltt, amsfonts, wrapfig, boxedminipage, color, cite, bookmark, slashbox} 
\usepackage[dvips]{graphicx}
\newcommand{\qed}{\nobreak \ifvmode \relax \else
      \ifdim\lastskip<1.5em \hskip-\lastskip
     \hskip1.5em plus0em minus0.5em \fi \nobreak
      \vrule height0.75em width0.5em depth0.25em\fi}

\newcommand{\eg}{{\it e.g., }}
\newcommand{\etal}{{\it et~al., }}
\newcommand{\ie}{{\it i.e., }}

\newcommand{\comments}[1]{}
\newcommand\hl{\bgroup\markoverwith\textsl{•}
  {\textcolor{yellow}{\rule[-.5ex]{2pt}{2.5ex}}}\ULon}

\newlength{\boxfigwidth}

\newcommand{\boxfig}[1]{
\begin{figure}[h]
\begin{center}
\begin{small}
\setlength{\boxfigwidth}{3.15in}
\addtolength{\boxfigwidth}{0in}
\noindent\framebox{\quad\begin{minipage}{\boxfigwidth}
#1
\vspace{-15pt}
\end{minipage}\quad}
\end{small}
\end{center}
\end{figure}
}

\begin{document}

\title{F-FDN: Federation of Fog Computing Systems for Low Latency Video Streaming}

\author{
    Vaughan Veillon, Chavit Denninnart, Mohsen Amini Salehi\\
	High Performance Cloud Computing (HPCC) Laboratory,\\ 
	School of Computing and Informatics, University of Louisiana at Lafayette, USA \\
    \{vcv5733, cxd9974, amini\}@louisiana.edu
    }


\maketitle

\IEEEpeerreviewmaketitle
\begin{abstract}
Video streaming is growing in popularity and has become the most bandwidth-consuming Internet service. As such, robust streaming in terms of low latency and uninterrupted streaming experience, particularly for viewers in distant areas, has become a challenge. The common practice to reduce latency is to pre-process multiple versions of each video and use Content Delivery Networks (CDN) to cache videos that are popular in a geographical area. However, with the fast-growing video repository sizes, caching video contents in multiple versions on each CDN is becoming inefficient. Accordingly, in this paper, we propose the architecture for Fog Delivery Networks (FDN) and provide methods to federate them (called F-FDN) to reduce video streaming latency. In addition to caching, FDNs have the ability to process videos in an on-demand manner. F-FDN leverages cached contents on the neighboring FDNs to further reduce latency. In particular, F-FDN is equipped with methods that aim at reducing latency through probabilistically evaluating the cost benefit of fetching video segments either from neighboring FDNs or by processing them. Experimental results against alternative streaming methods show that both on-demand processing and leveraging cached video segments on neighboring FDNs can remarkably reduce streaming latency (on average 52\%).
\end{abstract}


\section{Introduction}\label{sec:intro}

Video streaming occupies more than 75\% of the whole Internet bandwidth and it is predicted that the growth will persist~\cite{cdnstat}. 
The resources required to provide video streams is also increasing. High quality video (such as 4K) and interactive video streaming (such as 360 degree videos, story branching videos) are becoming commonplace. High quality streaming increases data rate consumption, while interactive video streaming demands low latency. 
With the demand for video content steadily increasing, the ability to effectively deliver video content to a viewerbase that is spread on a global scale is a major concern for video streaming providers~\cite{latval1}. 

To address increasing data rate concerns, streaming providers require large-scale computing and storage resources. Therefore, many video providers (\eg Netflix) have migrated to cloud services to host and deliver their video contents. Using cloud services alleviates the burden of maintaining and upgrading physical resources from the video streaming provider~\cite{salehi10}. For instance, since 2015, Netflix stopped using their own datacenters and moved their entire streaming service to Amazon cloud (AWS) and Open Connect Appliances (OCA)~\cite{netflix}. Also, YouTube utilizes Google cloud services to achieve their streaming demands~\cite{hoiles2015adaptive}. However, the latency of accessing cloud services can be significant, specifically, for viewers that are distant to the cloud servers~\cite{latval1}. In order to overcome this inherent latency issue, stream providers commonly utilize a distributed system known as a CDN~\cite{latval1}. A CDN caches part of the video repository into its edge locations that are physically close to viewers resulting in a lower latency compared to accessing from a more centrally located cloud server. 

The problem is that the large and fast-growing repository size of streaming providers has made it infeasible to cache a large portion of the overall content on their CDNs. In addition, caching on CDNs is less effective because of the fact that streaming providers have to maintain multiple versions of the same video to be able to support heterogeneous display devices and network conditions~\cite{cvss}. 
As such, instead of pre-processing video streams into multiple versions, mechanisms for on-demand processing (\eg on-demand transcoding~\cite{cvss}) of video streams is becoming prevalent~\cite{CVSSJournal,tpds18}. 
However, the challenge is that on-demand video processing cannot be performed on CDNs since they are predominantly used for caching purposes~\cite{cdninfo}.

These limitations lead to frequent streaming directly from central cloud servers, which increases streaming latency, hence, decreasing viewers' \emph{Quality of Experience} (QoE), particularly, in distant areas~\cite{khan2004performance}.
To overcome these limitations, in this research, we propose a novel distributed platform, named a \emph{Federated-Fog Delivery Network} (F-FDN), that leverages the computing ability of fog systems to carry out on-demand processing of videos at the edge level. F-FDN is composed of several Fog Delivery Networks (FDN) that collaboratively stream videos to viewers with the aim of minimizing video streaming latency.

Using F-FDN, video streaming providers only need to cache a base version of a video in an edge (fog) and process them to match the characteristics of the viewers' devices in an on-demand manner. 
More importantly, F-FDN can achieve location-aware caching (\ie pre-processing) of video streams. That is, video streams that are popular (\ie hot) in a certain region are pre-processed and cached only in that region. Due to resource limitations of FDN, we propose to pre-process only the hot portions of videos~\cite{hotness} and the remaining portions are processed in an on-demand manner. To alleviate the on-demand processing load in an FDN, we develop a method to leverage the distributed nature of F-FDN and reuse pre-processed video contents on neighboring FDNs. This allows the streaming of different portions of a video from multiple sources (\ie FDNs), subsequently, increasing viewers' QoE.

In summary, the contributions of this research are as follows:
\begin{itemize}
\item Proposing F-FDN platform to improve QoE for viewers' located in distant areas.
\item Developing a method within each FDN to achieve video streaming from multiple FDNs simultaneously.
\item Analyzing the impact of F-FDN on the viewers' QoE, under varying workload characteristics.
\end{itemize}

The rest of this paper proceeds as follows: Section~\ref{sec:bkgd} covers background information regarding video streaming tendencies. We provide an overview of our system in Section~\ref{sec:overview} with a detailed explanation of the different components of the F-FDN architecture. In Section~\ref{sec:seg} the decision making policies we have put in place to deliver a video stream will be further discussed. In Section~\ref{sec:implementation}, the alternative streaming methods we test against are explained, and in Section~\ref{sec:evltn} we explain the set up of the experiments and show the results. In Section~\ref{sec:rw} we cover other works related to this research. Finally in Section~\ref{sec:conclsn} we conclude the paper and mention our future considerations for this research.

\section{Background}\label{sec:bkgd}

Video streaming is achieved via processing and streaming independent video segments in the form of Group Of Pictures (GOP)~\cite{jokhio2011analysis}. A video stream generally consists of multiple GOPs, depending on the content type and length (\ie timespan) of the video~\cite{tpds18}.

QoE of the viewer is defined as the ability to stream each GOP within its allowed latency time to create an uninterrupted streaming experience. The allowed latency time for a GOP is its presentation time, hence, that is considered as the GOP's deadline~\cite{jokhio2011analysis,matin_paper}. 

A large body of research studies have been undertaken to maintain the desired video streaming QoE (\eg~\cite{cvss,CVSSJournal}) through efficient use of cloud services. 
Particularly, the earlier studies have shown that the access pattern of video streams follows a long-tail distribution~\cite{darwich16,hotness}. That is, only a small percent of videos are streamed frequently (known as \emph{hot} video streams) and the majority of videos are rarely streamed~\cite{miranda}. For instance, in the case of YouTube, it has been shown that only 5\% of videos are hot~\cite{gill2007youtube}. 
It has also been shown that, even within a hot video, some portions are accessed more often than others. For instance, the beginning portion of a video or a popular highlight in a video are typically streamed more often than the rest of the video~\cite{hotness}. 

Considering this long-tail access pattern, streaming service providers commonly pre-process (store) hot videos or GOPs in multiple versions to fit heterogeneous viewers' devices~\cite{hotness}. Alternatively, they only keep a minimal number of versions for the rarely accessed videos~\cite{cvss,vlsc}. For any portions of the video that are not pre-processed, they are processed in an on-demand manner upon viewer's request~\cite{CVSSJournal}.

A video stream of an interactive video, such as 360 degree or story branching videos, changes based on how a viewer is consuming it. These types of videos benefit greatly from lowered streaming latency. In general, if the latency is high, the viewer will need a higher amount of buffer to cover processing and streaming delay. Based on the nature of interactive videos, some parts of the buffer may end up not being viewed. Therefore, low latency streaming reduces the amount of buffer needed and thus reduces the amount of wasted processing.


\begin{figure}
  \centering
  \includegraphics[width=0.5\textwidth]{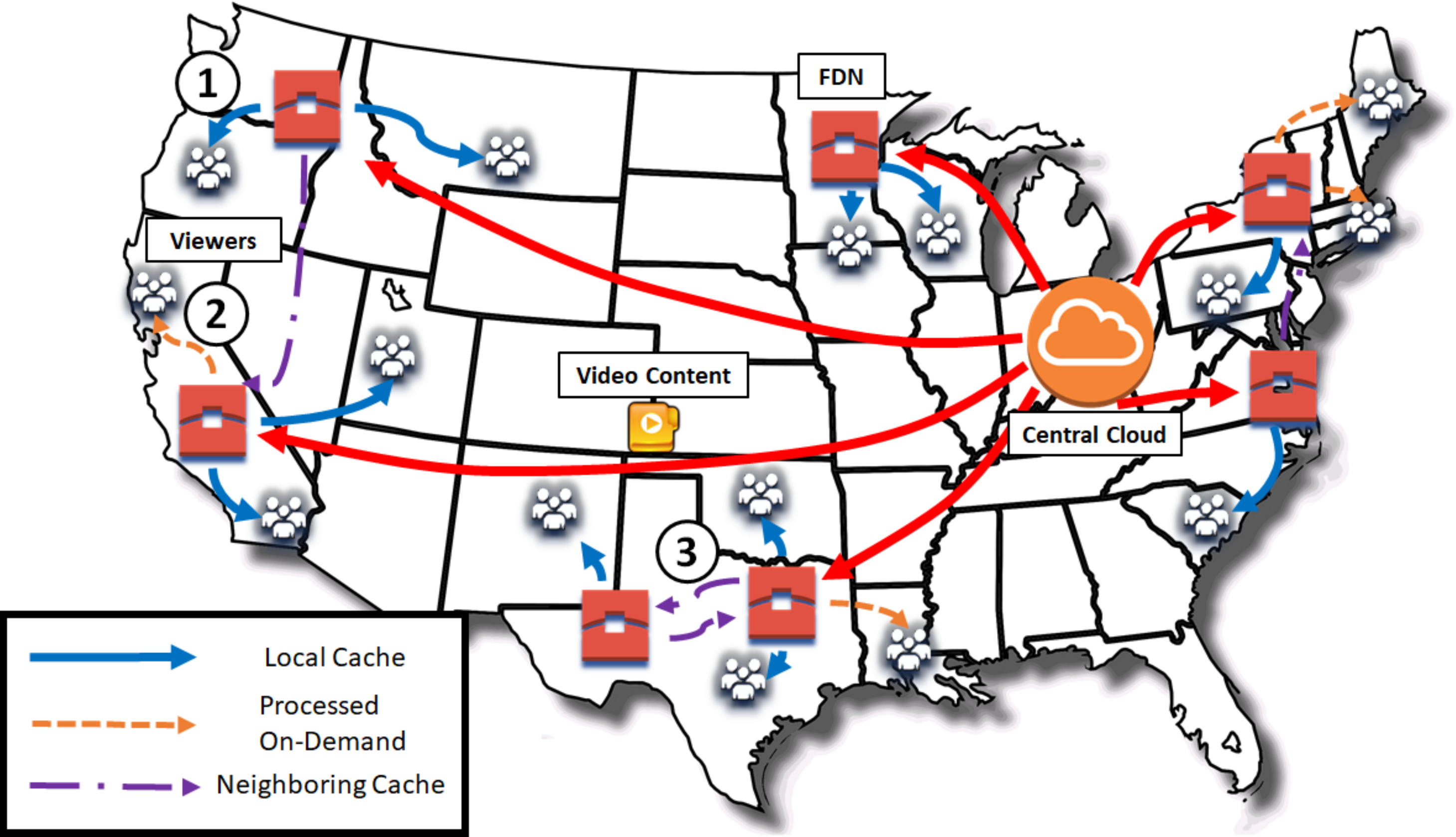}
  \caption{High level view of the F-FDN architecture. Note the viewers that are receiving video content from multiple FDN. 1) shows video content coming from FDN local cache 2) shows video content being processed on-demand 3) shows video content coming from a neighboring FDN's cache}
  \label{fig:arch}
\end{figure}

\section{Federated Fog Delivery Networks (F-FDN)}\label{sec:overview}
\subsection{Overview}\label{sec:agt}

\begin{figure*}[ht]
  \centering
  \includegraphics[width=0.8\textwidth]{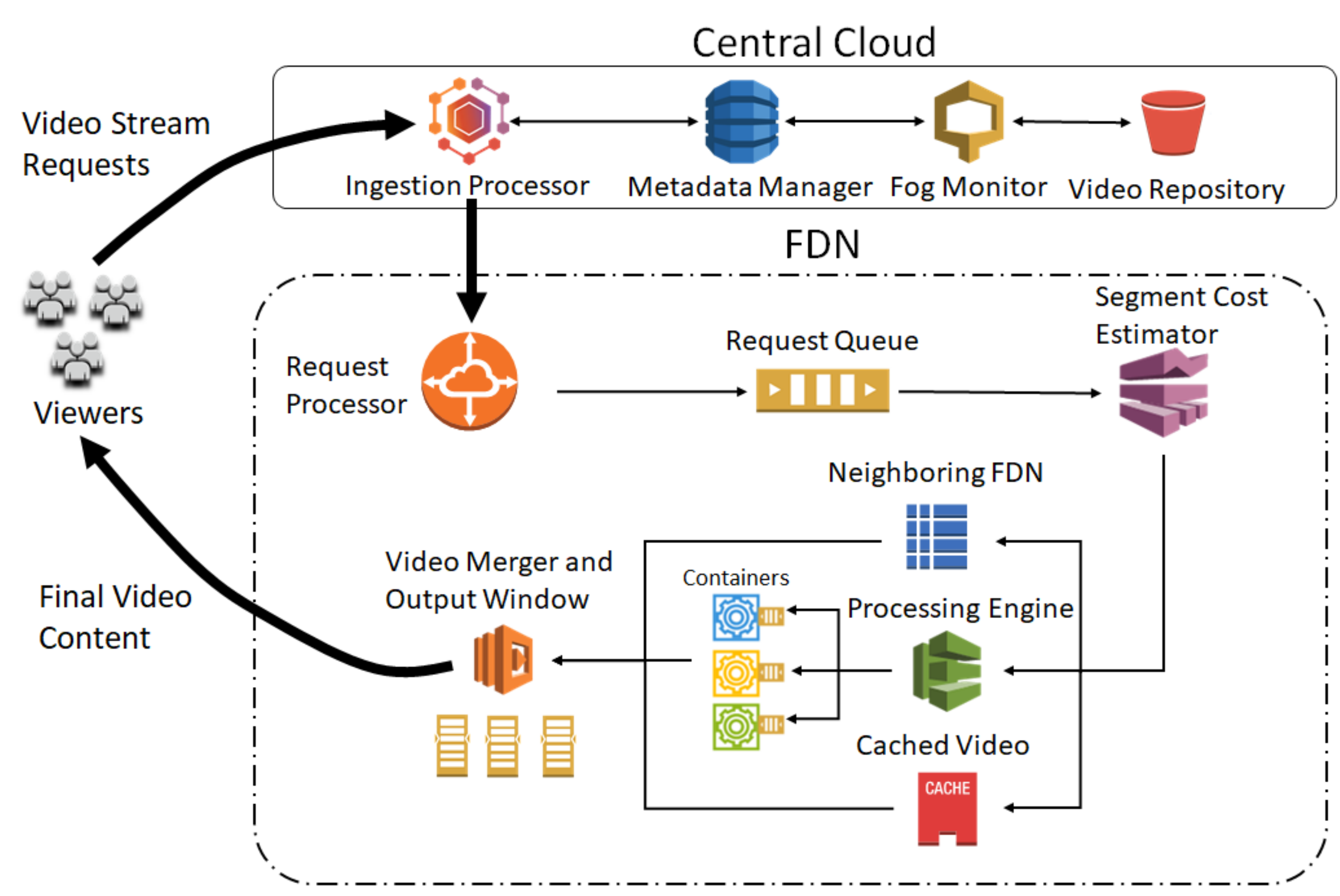}
  \caption{System Components of the F-FDN architecture. It is composed of a centrally located cloud and a federation of multiple Fog Delivery Networks (FDNs).}
  \label{fig:components}
\end{figure*}

The aim of F-FDN is to deliver the highest possible QoE to viewers, independent of their geographical location. We position F-FDN to achieve this by utilizing all of the resources that we have available in the system. One of the most differentiating qualities of F-FDN compared to other video streaming systems is the ability to evaluate on a segment by segment basis how to stream a video. It is worth noting that, in traditional CDNs, the entire video is always streamed from the same CDN server as long as a connection is maintained with the viewer~\cite{stocker2017}.

F-FDN is composed of several connected, peer Fog Delivery Networks (FDNs) that are also connected to a central cloud server. An FDN caches pre-processed GOPs for videos that are hot in that region. This results in each FDN having varied pre-processed video content that is optimized to the viewers local to that FDN.
When a video stream is requested, the viewer is connected to its most local FDN. As the video is being streamed, decisions are made on a segment (GOP) by segment basis as to how the GOP is delivered to the viewer, so that the likelihood of meeting the GOP's deadline is maximized. The process in making these decisions is described in detail in Section~\ref{sec:seg}. 
 
An example of a video being streamed to a viewer from multiple sources can be observed in Figure~\ref{fig:arch}. As we can see in the figure, using F-FDN, a video segment can be streamed to a viewer in three different ways:

\begin{enumerate}
\item FDN Local Cache: the FDN local to the viewer already has the requested video segment in it's cache and streams the segment to the viewer.
\item Processed On-demand:
the local FDN processes the missing (\ie non-cached) segments according to the characteristics of the request and then stream them to the viewer.
\item Neighboring FDN's Cache: 
the missing segments exist in the cache of a neighboring FDN, the segment is then transferred to the local FDN and then streamed to the viewer.
\end{enumerate}

At a high level, F-FDN is composed of two main components, namely a Central Cloud and a distributed network of FDNs. Figure~\ref{fig:components} shows the internals of the Central Cloud and each FDN. Details of each component is elaborated in the next subsections.

\subsection{Central Cloud}
The Central Cloud, with virtually unlimited storage and processing capabilities, is where all streaming requests are initially ingested and where all the FDNs in the system are managed. As we can see in Figure~\ref{fig:components}, Central Cloud consists of four main components:

\begin{itemize}
\item Ingestion Processor - handles all the incoming stream requests made by viewers. It determines which FDN is the most local to the viewer in order to start the video streaming process. The FDN that is selected is the one that determines the way each video segment is streamed to the viewer.
\item Metadata Manager - keeps track of all the cached video segments contained throughout the FDNs. In addition, it keeps track of other metadata, such as file size and network latency between neighboring FDNs. In Section~\ref{sec:seg}, we explain how this metadata helps in determining the way to stream a video to the viewer. 
\item Fog Monitor - keeps track of the availability of FDNs via sending heartbeat pings to them. Also, it evaluates network latency between FDNs which is then communicated to the Metadata Manager to maintain up-to-date information.
\item Video Repository - contains a repository of all videos, pre-processed in multiple versions. This is where any cached video content on FDNs originates from. In the event that one or more video segments are missing in a local FDN, one decision can be fetching the segments from the video repository of the Central Cloud.
\end{itemize}

\subsection{Fog Delivery Network (FDN)}
Each FDN consists of six components, as shown in Figure~\ref{fig:components} and explained below.
\begin{itemize}
\item Request Processor - receives video stream requests from the Central Cloud and put them into a request queue. 

\item Segment Cost Estimator - makes the decision as to how each video segment in a video stream request should be streamed to viewers with minimum latency (\ie maximum likelihood of meeting the segment's deadline). For a given video segment, it is determined whether the segment should be streamed from the local FDN's cache, processed on-demand by the local FDN, or retrieved from a neighboring FDN's cache and then streamed by the local FDN. The process by which these decisions are made is explained in Section~\ref{sec:seg}.

\item Neighboring FDN Metadata - contains knowledge of all cached video segments on other FDNs and the network latency for accessing them. It is worth noting that the latency of accessing Central Cloud is also maintained by this component.

\item On-demand Processing Engine - is in charge of on-demand processing of video streams. We have developed the engine in our prior studies~\cite{cvss,CVSSJournal}. The engine uses multiple worker Virtual Machines (VMs) and enable the FDN to process incoming video segments based on the characteristics of the viewer's device.     

\item Cached Video Segments - GOPs of videos that are determined to be hot~\cite{hotness} in a region are pre-processed and cached by the FDN. For a given video streaming request, if some of its segments are locally cached, they impose the minimum network latency and cause higher QoE for the viewer.

\item Video Merger and Output Window - where the video segments of the video stream are put in the correct order and then streamed to the viewer.
\end{itemize}

\section{Maximizing Robustness of F-FDN}\label{sec:seg}

Streaming service providers aim at providing an uninterruptable streaming experience to their viewers. They strive to avoid and minimize missing deadlines of streaming tasks. The distributed nature of F-FDN provides multiple options (sources) to stream a single video segment. To minimize missing presentation deadline of a video segment, it should be streamed from the source that imposes the minimum streaming latency, hence, offering the maximum probability to meet the segment's deadline.

The streaming latency is affected by two main factors, namely video segment processing time and transfer time across the network. In particular, both of these factors have a stochastic nature~\cite{stochcloud}. An ideal method to stream videos in F-FDN should be robust against these stochasticities. That is, the method should maintain its performance, in terms of meeting the deadlines of streaming tasks, even in the presence of these uncertainties.
We implement a method within FDNs to account for the uncertainties of F-FDN and maximize the probability of meeting deadline for a given streaming task. This method is utilized within the Segment Cost Estimator component of FDN and makes the FDN robust.

\begin{table}[ht]
\centering
\caption{Important symbols used in Section~\ref{sec:seg}.}
\label{tb:tablename}
\begin{tabular}{l|l}

\textbf{Symbol} & \textbf{Description}\\
\hline
\hline
\textbf{\small{$s_i$}}		 & size of video segment $i$\\
\hline
\textbf{\small{$r_i$}}       & probability of processing segment $i$ \\ & on time (robustness)\\           
\hline
\textbf{\small{ $N_i^C(\mu_i,\sigma_i$)}}     & overall delivery distribution \\ & (end-to-end latency) for segment $i$  \\
\hline
\textbf{\small{$N_i^\tau(\mu_{jv},\sigma_{jv}$)}} & distribution of network throughput between\\ & two points for a given segment $i$\\
\hline
\textbf{\small{$N_{i}^E(\mu_{ij},\sigma_{ij})$}}    & processing time distribution for segment $i$ \\

\end{tabular}
\end{table}

\subsection{Network Latency of Streaming a Video Segment in FDN}
For a given video segment $i$, the latency probability distribution of transferring it between two points can be obtained based on the segment size (detnoted $s_i$) and the amount of data that can be transferred within a time unit between the two points (\ie network throughput). Prior studies show that the latency probability distribution to transfer video segment $i$ follows a normal distribution (denoted $N_i^\tau$)~\cite{normal,tpds18}. The two points can be between two FDNs or an FDN and a viewer. 


\subsection{Robust Video Segment Delivery in F-FDN}
We formally define $robustness$ of segment $i$, denoted $r_i$, as the probability that segment $i$ is delivered to the viewer's device before or at its deadline $\delta_i$. 

As mentioned earlier, each video segment can be retrieved using one of the following choices: (A) from the FDN's local cache; (B) processing it on-demand in the local FDN; (C) from a neighboring FDN's cache. 

In choice (A), the robustness of delivering segment $i$ is obtained from the segment latency probability distribution between local FDN $j$ and viewer's device $v$.
As such, the probability distribution for delivering a segment to the viewer, denoted $N_i^C(\mu_i,\sigma_i)$, for choice (A) is determined using Equation~\ref{eq:choiceA}.  

\begin{equation}\label{eq:choiceA}
 N_i^C(\mu_i,\sigma_i)= N_i^\tau(\mu_{jv},\sigma_{jv})
\end{equation}

In choice (B), the latency is impacted not only by the segment latency probability distribution between FDN $j$ and the viewer's device $v$, but also by the time to process the segment in FDN $j$. The processing times of a video segment can be estimated based on a probability distribution. This distribution is obtained from historical execution times of a particular processing type (\eg bit-rate transcoding) for a segment. It has been shown that the processing time of a video segment exhibits a normal distribution~\cite{tpds18}. Let $N_{i}^E(\mu_{ij},\sigma_{ij})$ be the probability distribution of completing the processing of segment $i$ on FDN $j$; also let $N_i^T(\mu_{jv},\sigma_{jv})$ be a normal distribution representing latency to deliver segment $i$ from the local FDN $j$ to the viewer's device $v$. Then, the probability distribution of delivering segment $i$ to the viewer is calculated by convolving the two distributions as shown in Equation~\ref{eq:choiceB}.

\begin{equation}\label{eq:choiceB}
 N_i^C(\mu_i,\sigma_i)=N_{i}^E(\mu_{ij},\sigma_{ij}) * N_i^\tau(\mu_{jv},\sigma_{jv})
\end{equation}

Similarly, the latency for choice (C) is impacted by two factors, the latency distribution for retrieving a segment from a neighboring FDN $k$ to the local FDN $j$, denoted $N_{i}^T(\mu_{kj},\sigma_{kj})$, and the segment latency distribution between FDN $j$ and viewer's device $v$, denoted $N_i^\tau(\mu_{jv},\sigma_{jv})$. 
Therefore, to obtain the probability distribution of delivering segment $i$, we convolve these two probability distributions as shown in Equation~\ref{eq:choiceC}.

\begin{equation}\label{eq:choiceC}
N_i^C(\mu_i,\sigma_i)=N_{i}^\tau(\mu_{kj},\sigma_{kj}) * N_i^\tau(\mu_{jv},\sigma_{jv})
\end{equation}

Once we have the final distribution, $N_i^C(\mu_i,\sigma_i)$ the robustness of segment $i$ can be measured using its deadline $\delta_i$ based on Equation~\ref{eq:prob}. In fact, in this case the liklihood that segment $i$ can be delivered before $\delta_i$ is the cumulative probability for a random variable $X$ to be less than or equal to $\delta_i$ which is the robustness of segment $i$.

\begin{equation}\label{eq:prob}
r_i=P(X \leq \delta_i)
\end{equation}

The algorithm for the Segment Cost Estimator (shown in Figure \ref{fig:algdrop}) utilizes the robustness for each segment of a video stream to determine how to fetch that segment, hence, assuring a high quality and uninterruptable video streaming experience for viewers. The algorithm first checks if segment $i$ exists in the local cache of FDN $j$ to be streamed to the viewer (Step 1). If it does not exist locally, then a list of all neighboring FDN containing segment $i$ is retrieved from the Metadata Manager and their respective robustness values are calculated (Steps 2---7). The robustness of on-demand processing segment $i$ is also calculated and compared against the neighboring FDN that has segment $i$ with the highest robustness (Steps 8---9). Finally, in Step 10, the option with the highest robustness is chosen to provide segment $i$.

\boxfig{





Upon receiving a video stream request $m$, at $FDN_j$: 

For every segment $i$ in video stream request $m$:
	\begin{itemize}
	\item[(1)] if $i$ exists in $FDN_j$'s local cache, stream $i$ to viewer
	\item[(2)] else if $i$ is available in neighboring FDNs or in central cloud:
		
	\begin{itemize}
	\item[(3)] retrieve list of metadata of all remote locations that match segment $i$
	\item[(4)] For each metadata item $l$ in the metadata list:
		\begin{itemize}		
		\item[(5)] calculate the cumulative probability of the transfer time from $l$ using the presentation time of $i$
		\item[(6)] track which FDN $l$ offers the highest probability of success		
		\end{itemize}		
	\item[(7)] convolve processing and transfer time distributions for segment $i$ in $FDN_j$ and calculate its cumulative probability
	\item[(8)] compare the probability of processing $i$ on-demand with the probability of streaming $i$ from FDN offers the highest probability of success
	\item[(9)] stream segment $i$ from the option with the highest probability of satisfying $i's$ deadline			
	\end{itemize}
\end{itemize}
\caption{Procedure followed by Segment Cost Estimator} \label{fig:algdrop}
}

\section{Methods for Video Streaming Delivery}
\label{sec:implementation}


\begin{table*}[ht]
\centering
\caption{Characteristics of various methods implemented to examine the performance of the F-FDN platform.}
\label{tb:tablename}
\begin{tabular}{l|l|l|l|l}
\backslashbox{\textbf{\small{Methods}}}{\textbf{\small{Characteristics}}} & \textbf{\parbox{1.5cm}{\small{Caching at \\ Edge}}} & \textbf{\small{Federated}} & \textbf{\parbox{1.8cm}{\small{On-demand \\ Processing}}} & \textbf{\parbox{2cm}{\small{Robustness \\Consideration}}}  \\
\hline
\hline
\textbf{\small{Central Cloud}}     & no              & no               & no              & no                 \\
\hline
\textbf{\small{CDN}}               & yes             & no               & no              & no                 \\
\hline
\textbf{\small{Federated CDN (F-CDN)}}    & yes             & yes              & no              & yes                \\
\hline
\textbf{\small{Isolated FDN (I-FDN)}}      & yes             & no               & yes             & yes                \\
\hline
\textbf{\small{Deterministic F-FDN}} & yes             & yes              & yes             & no                 \\
\hline
\textbf{\small{Robust F-FDN}}       & yes             & yes              & yes             & yes    \\

\end{tabular}
\end{table*}

In this section, we explain alternative methods for stream delivery. Table~\ref{tb:tablename} provides an overview for the various methods we implemented and highlights differences in their characteristics. These methods encompass current practices for video streaming (namely, CDN and Central Cloud) and baseline methods (namely, F-CDN, Isolated FDN, and Deterministic F-FDN) that focus on various aspects of the F-FDN platform in isolation. Finally, the \emph{Robust F-FDN} is the streaming delivery method operating based on the theory developed in Section~\ref{sec:seg}. It is noteworthy that these methods are implemented within the \emph{Segment Cost Estimator} component of the FDN (see Figure~\ref{fig:components}). 
The rest of this section further elaborates on the characteristics of the implemented methods that are used in the experiment section.
\medskip

\noindent\textbf{Central Cloud.}
This method considers only the central cloud where all the video contents are available in the main video repository. Every video segment is streamed directly from the cloud and no geographically spread FDN or CDN are considered to reduce the streaming latency.

\medskip

\noindent\textbf{CDN.} Due to popularity of the CDN approach in the streaming industry, we consider it in our evaluations. Our simulated CDN consists of a central cloud that holds the same characteristics as the previously described system and CDN servers which have 75\% of the requested videos cached, which is a realistic level for CDN caching ~\cite{cache2016}. As CDNs are located close to viewers, any segments streamed from them have a lower latency compared to the central cloud. It is noteworthy that CDN servers do not perform any computation and  caches the entirety of a video, rather than only a portion. Also, any segment that is not found in a CDN, is streamed from the central cloud.

\medskip

\noindent\textbf{Federated CDN.} The Federated CDN (F-CDN) includes a central cloud and CDN servers in its system. The key difference of F-CDN with CDN is partial video caching. In F-CDN, it is possible to cache only few segments of a video, rather than the entirety of the video. Owing to the federated nature, in F-CDN, video segments can be streamed from the local CDN, a neighboring CDN, or from the central cloud. The rationale of implementing this method is to study the impact of federation of cached contents, without the ability to process videos on-demand. This method makes use of the robustness definition, introduced in Section~\ref{sec:seg}, to stream a given video segment from the CDN that offers the highest probability to meet the deadline of that segment.

\medskip

\noindent\textbf{Isolated FDN (I-FDN).} The Isolated FDN method includes a central cloud and a single FDN. In this system, the FDN node performs on-demand processing of video segments, in addition to caching. However, it does not consider retrieving segments from neighboring FDNs. That is, the segments are streamed only from the FDN's cache, processed on-demand, or from the central cloud. The streaming decisions for each video segment is made between the local FDN and central cloud based on the robustness definition (see Section~\ref{sec:seg}). The rationale of implementing this method is to study the impact of lack of federation on the streaming QoE.

\medskip

\noindent\textbf{Deterministic F-FDN.} 
The Deterministic F-FDN method consists of a central cloud and a federation of FDNs. While each FDN can perform caching and on-demand processing, the federation enables the option to stream cached segments from neighboring FDNs as well. In the Deterministic F-FDN, for each segment, streaming decisions are only made based on expected transmission and processing times, \ie it ignores the stochastic nature that exists in the F-FDN environment. This method demonstrates the impact of ignoring uncertainties that exist in the system on the overall streaming QoE.

\medskip

\noindent\textbf{Robust F-FDN.} 
Unlike Deterministic F-FDN, the Robust F-FDN operation takes into account the stochastic nature that exists in both communication and computation of the F-FDN platform. The more informed decision making is expected to have more streaming tasks meeting their deadlines, resulting in a more robust streaming service, regardless of the viewers' geographical location. It is, in fact, the implementation of the theory developed in Section~\ref{sec:seg} and the method described in the algorithm of Figure~\ref{fig:algdrop}. 

\section{Performance Evaluation}
\label{sec:evltn}

\section*{Experimental Setup}

We conducted an emulation study to understand the behavior of the F-FDN platform. We developed a prototype of F-FDN by expanding the CVSS (Cloud Video Streaming Service) platform~\cite{CVSSJournal,cvss,chavit}. The prototype has the ability to operate for all video streaming methods described in Section~\ref{sec:implementation}. Within CVSS, we simulated three worker VMs that are modeled after the Amazon GPU (\texttt{g2.2xlarge}) VM to perform the video processing. The reason we considered GPU-based VMs is that in~\cite{tpds18} it is shown that these VM types fit the best for video processing tasks. Our experiments are conducted by using three FDNs in the system, in addition to a central cloud server. Our experiments consider streaming requests arriving to one of these FDNs and we measure the performance metrics obtained in that FDN. The other two FDNs serve as neighbors, caching a portion of video segments (as detailed in Section~\ref{subsec:cach}).

We generated different workload traces of video stream requests to examine behavior of the system under various workload conditions and streaming methods. The workloads used in the experiments are created using a set of benchmark videos that contain different lengths and content types. The benchmark videos are publicly available for reproducibility purposes at \url{https://goo.gl/TE5iJ5}. For each video segment in the workload traces, there is an associated processing (\ie execution) time, which is obtained from the mean of 30 times execution of that segment on Amazon GPU VM. For the sake of accuracy and to remove uncertainty in the results, we generated 30 different workload traces. Each workload trace simulates 3 minutes
of video stream request arrivals. The arrival time of each streaming request is determined by uniformly sampling within the time period. All segments of the same video have the same arrival time but different deadlines (\ie presentation times). Accordingly, each experiment is conducted with the 30 workload traces and the mean and 95\% confidence interval of the results are reported.

We track the number of deadlines that are missed, which indicates the robustness of the system. A deadline is considered missed due to a segment being streamed after its associated presentation time. The presentation time of a segment is based on the order of the segment's appearance in a video. As we consider a Video On-demand streaming service, even if a segment misses its deadline, it still must complete its execution and is streamed to the viewer.

To consider bandwidth usage in the evaluations, we have a limited bandwidth value from the local FDN to the viewer, and from the local FDN to other neighboring FDN. This bandwidth value becomes more congested as segments are initially streamed and less congested as segments finish streaming. Each node also has an associated latency value. The network latency values used for the edge servers and the central cloud server were taken from \cite{latval1,latval2}. For our bandwidth values, we used an average of 1 Gbps.

\section*{Experimental Results}

\subsection{Analyzing Suitable Cache Size for FDNs}\label{subsec:cach}
\begin{figure}
  \centering
  \includegraphics[width=0.5\textwidth]{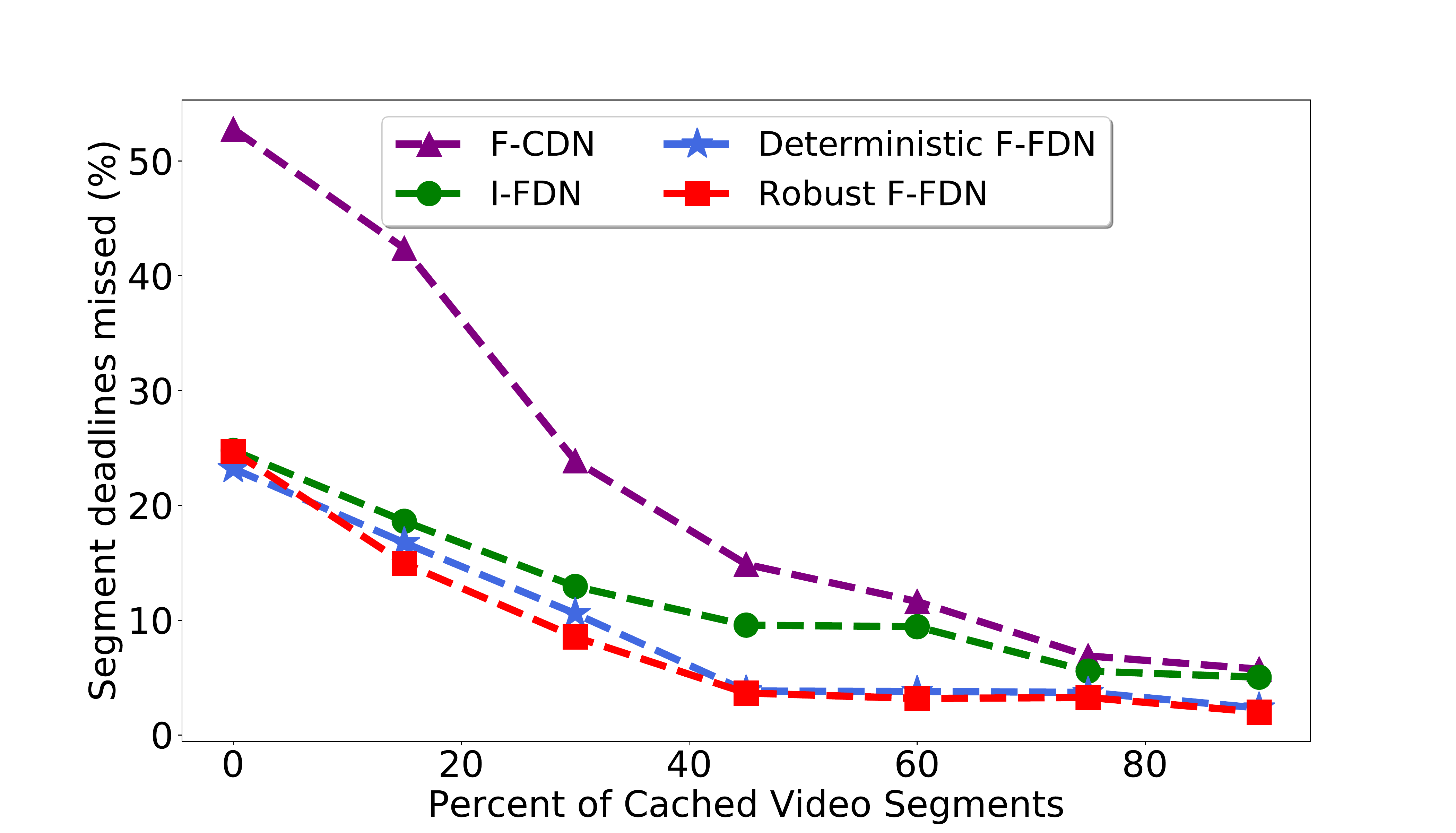}
  \caption{Deadline miss rate of different streaming methods as the caching level is increased. The simulations are run using a work load of 3500 segments.}
  \label{fig:cache}
\end{figure}

In our first experiment, we intend to find the minimum percentage of video contents that needs be cached within an FDN, so that a high level of QoE is maintained for viewers. Recall that we measure QoE in terms of percentage of video segments missing their presentation deadline. We evaluate variations of FDNs (namely, F-CDN, I-FDN, Deterministic F-FDN, and Robust F-FDN) to understand how different methods take advantage of the caching feature. For that purpose, as shown in Figure~\ref{fig:cache}, we increase the percentage of video segments that are cached in each FDN (horizontal axis) and measure the percentage of video segments that miss their deadlines (vertical axis). In this experiment, we used workload traces consisting of 3,500 segments being streamed to viewers. 

We observe that as the percentage of cached segments is increased, the deadline miss rate drops remarkably across all methods---from approximately 53\% to around 2\%.
Specifically, when the total cached video content is at zero percent, we are able to see a major difference between F-CDN and the other three systems. Zero percent caching for F-CDN, in fact, shows the case of streaming only from the central cloud. However, we can observe that other methods with the ability to process segments at the fog (FDN) level, in addition to streaming segments from cloud, can dramatically reduce deadline miss rate (approximately 52\% improvement). 

As the level of cached video content is increased, we observe the benefit of streaming video segments from neighboring FDNs. For instance, comparing the I-FDN and Deterministic F-FDN shows that at 30\% caching, deadline miss rate is reduced by 2.3\% (denoting 18\% improvement), whereas at 90\% caching, the deadline miss rate of Deterministic F-FDN is 2.7\% lower than I-FDN (denoting 53\% improvement). We can conclude that the streaming of video segments from neighboring FDNs unburdens the on-demand processing of the local FDN to the point where missing a deadline becomes significantly less likely.

Based on our observation and analysis in this experiment, we choose to use a caching level of 30\% for the FDN systems in the next experiments. We believe that this caching level provides a sustainable trade-off between caching size and streaming QoE.  

\subsection{Analyzing the Impact of Oversubscription}
In this experiment, our goal is to study the robustness of the F-FDN platform against increasing workload intensity (aka oversubscription) and compare it against alternative methods. For that purpose, we vary the number of arriving video segments from 3,000 to 4,500 (with increments of 500) within the same time interval and measure the percentage of video segments that miss their deadlines. In this experiment, FDN-based methods cache 30\% and the CDN method caches 75\% (for practical reasons~\cite{cache2016}) of video segments, while Central Cloud stores all the video contents.

\begin{figure}
  \centering
  \includegraphics[width=0.45\textwidth]{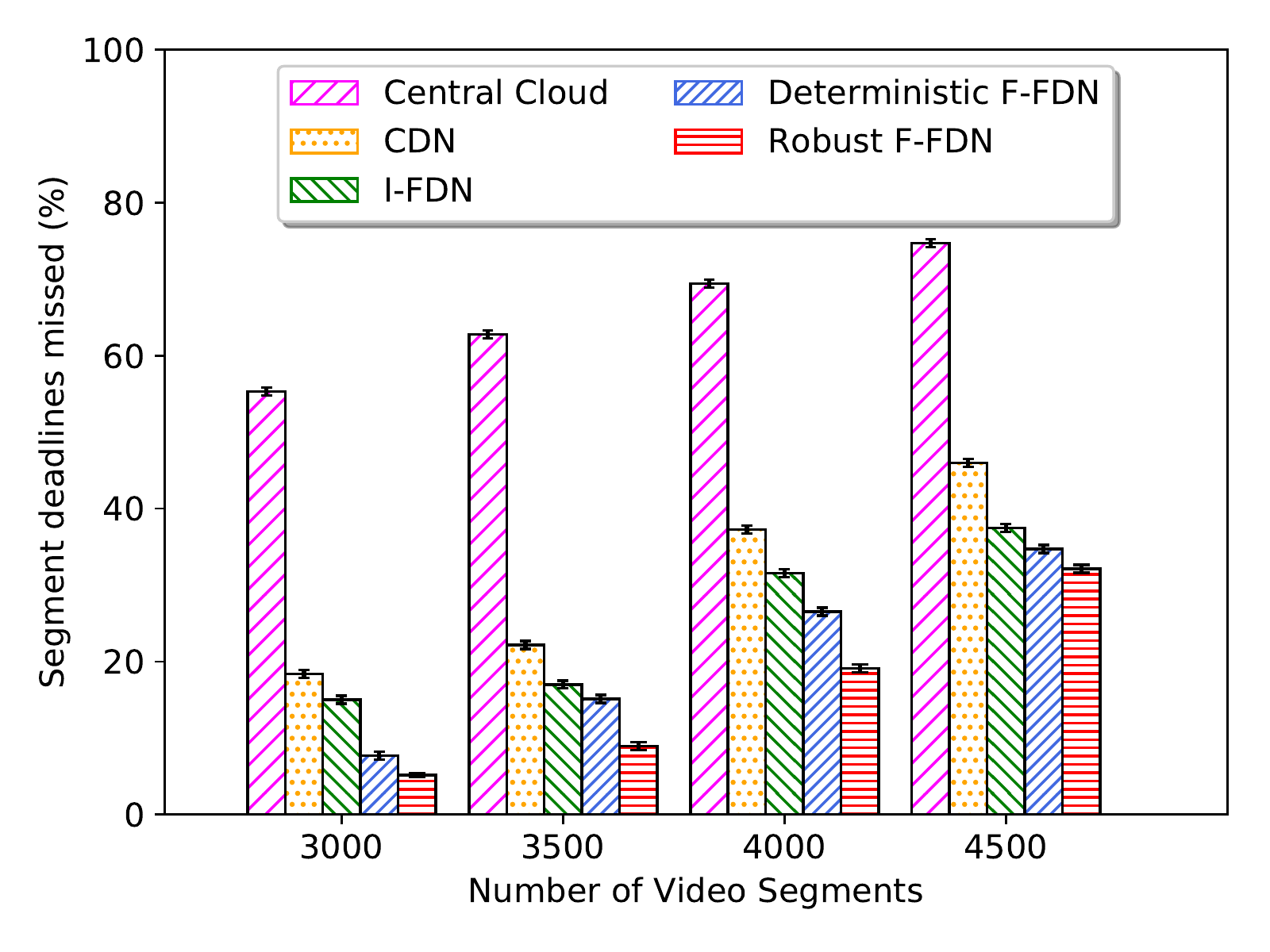}
  \caption{Deadline miss rate at increasing workload intensity. FDN-based  methods cache 30\% of video segments while CDN has 75\% of videos cached.}
  \label{fig:all}
\end{figure}

Figure~\ref{fig:all} demonstrates the performance of different methods as the workload size increases (horizontal axis). We observe that as the number of arriving requests increases, the percentage of segments missing their deadlines increases too. In particular, in comparing the CDN and Central Cloud methods, we see the benefit of a viewer being able to access a CDN server that is much closer to them geographically. Across all workloads the CDN method misses an average of 54\% less deadlines than the Central Cloud method. With the presence of on-demand processing in the I-FDN compared to CDN, there is an average of a 17\% deadline miss rate improvement. Performance is shown to further increase upon adding the federation of FDNs for streaming, as is present in the Deterministic and Robust F-FDN methods. Compared to the I-FDN, the Deterministic F-FDN has an average of 34\% less deadlines missed. 

Comparing the performance of the Deterministic F-FDN and the Robust F-FDN methods, we observe a further improvement of deadline miss rate. Across all workloads, the Robust F-FDN performs an average of 28\% better than the Deterministic F-FDN. This is due to capturing the stochastic factors (related to communication and computation) present in Robust F-FDN and absent in the Deterministic F-FDN. 

\subsection{Analyzing the Impact of Network Latency}

The goal of this experiment is to evaluate the robustness of our methods against viewers' geographical locations. This is particularly important for viewers located in distant areas, where the quality of the edge network commonly fluctuates and is highly uncertain. For that purpose, we study the performance of different methods where the uncertainty in the network latency between the viewer and FDNs and between FDNs is steadily increased.

\begin{figure}
  \centering
  \includegraphics[width=0.5\textwidth]{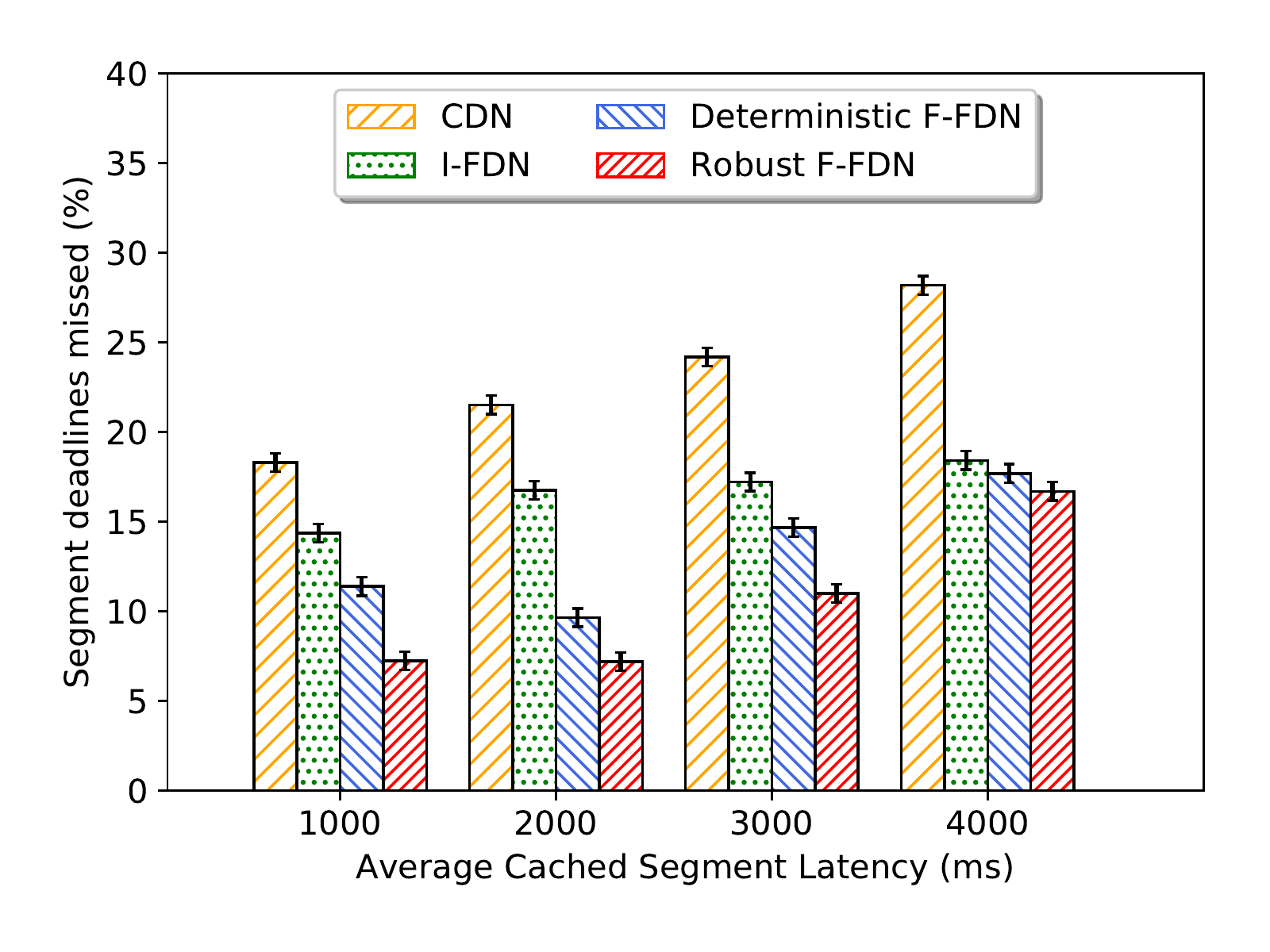}
  \caption{Deadline miss rate with increasing latency of the edge network. The experiments are conducted using 3500 video segments. FDN-based methods contain 30\% and CDN has 75\% cached video contents.}
  \label{fig:latency}
\end{figure}

The result of this experiment is shown in Figure \ref{fig:latency}. The horizontal axis shows the average latency to receive a cached video segment and the vertical axis shows the percentage of segments that missed their deadlines. We evaluated CDN, I-FDN, Deterministic F-FDN, and Robust F-FDN methods. Because our focus is on the edge network, we do not consider the Central Cloud method. 
This experiment is conducted with 3,500 segments and 30\% of video segments are cached in each FDN, except CDN that caches 75\% of all videos.

We observe that all methods result in a higher deadline miss rate as the network latency is increased. However, we observe that CDN deadline miss rate is increased at a greater rate than that of the other methods. When comparing the CDN and I-FDN system at an average network latency of 1,000 ms the deadline miss rate is at a difference of 20.2\%. Nonetheless, when the average network latency is increased to 4,000 ms, the difference in deadline misses maintains at 20.4\%. 

For the CDN method, all segments that are not cached must be streamed from the central cloud. When the latency for streaming from the CDN is not ideal, more segments are streamed from central cloud. Since I-FDN can perform on-demand processing for segments that are not cached, there is less of a reliance on the central cloud. This explains the consistently better performance of I-FDN as the average network latency is increased when compared to the CDN method.

In comparing I-FDN to Deterministic F-FDN, we notice that the difference of deadline miss rate decreases between the two methods as the network latency increases. Particularly, at an 1,000 ms network latency, the Deterministic F-FDN performs 64\% better than I-FDN. However, when the network latency is increased to 4,000 ms, there is only a 7.6\% difference in deadline miss rate between the two methods. This decrease in difference of deadline miss rates can be explained by the Deterministic F-FDN choosing to stream segments from neighboring FDNs less, instead, processing those segments in an on-demand manner. The same phenomenon can be observed when comparing I-FDN and Robust F-FDN. At 1,000 ms network latency, the deadline miss rate difference is 97\% and at 4,000 ms network latency, the difference is only 15.4\%. This observation shows the adaptability that is inherent to our system. 

In general, we observe that Robust F-FDN outperforms other methods. However, the difference between Deterministic F-FDN and Robust F-FDN becomes less prominent, when the network latency at the edge is at the highest we tested (4,000 ms). The reason for outperformance is capturing stochasticity in the network latency. However, the reason for similar performance at 4,000 latency is that both methods choose to process on-demand as opposed to relying on a highly uncertain network and fetch segments from neighboring FDNs.

We can conclude that F-FDN platform, in general, can remarkably improve the performance of streaming compared to traditional CDN-based methods. Interestingly, the improvement becomes more significant, as the network becomes more uncertain. Among FDN-based methods, Robust F-FDN can capture the stochasticity that exists in the network to a certain level and further improve the performance.

\section{Related Works}\label{sec:rw}

With traditional CDNs, the CDN servers are only used to cache data. This implies that any update to the CDNs' contents is dependent on centrally located cloud servers. With the integration of fog/edge computing into a distributed system like a CDN, computation can be performed on the network edge, near data users. Having the computation performed closer to the data source reduces the streaming latency, hence, higher QoE~\cite{edgepromise}. Alternatively, our system utilizes fog computing to perform on-demand video processing to reduce video streaming latency.

Li \etal~\cite{cvss} developed an architecture for CVSS, Cloud-based Video Streaming Service. CVSS utilizes cloud resources to deliver video streams through a balanced combination of on-demand processing and partial caching in order to minimize use of resources and maintain high QoE to viewers. Our work integrates CVSS into a distributed system in the form of F-FDN, where CVSS is used within each FDN.  

Lin \etal~\cite{cloudfog} propose a system, called CloudFog, that utilizes fog computing to enable thin-client Massive Multiplayer Online Gaming while maintaining a high user QoE. Their system works by having powerful and centrally located servers perform the computational tasks that are associated with the game state. The less computationally intensive task of rendering and streaming game video is handled by intermediate machines (called supernodes) that are physically closer to the users. This enables a user to play the game without the need of a powerful device, since the heavy computation is handled within the system enabled by fog computing. Similar to~\cite{cloudfog} we stream video from physically close servers (\ie FDN). However, our work is different than~\cite{cloudfog} in the sense that our system has more intelligence in reusing contents on peering FDNs. This enables F-FDN to operate with a greater independence from the central cloud. 

Ryden \etal~\cite{nebula} provides an architecture for a fog system, called Nebula, designed for applications where user data is geographically spread. An example of this type of application is managing video feeds from multiple cameras spread amongst various locations. Nebula specializes in performing location-aware and and location-specific processing of data-intensive and compute-intensive tasks. They were able to fully utilize the fog machines by monitoring the data storage and computational potential of the machines by forming machine groups based on their proximity to neighboring machines. Their methodology allows for multiple machines to compute or store data that is only relevant to its location. In a similar fashion, F-FDN keeps updated knowledge of the data stored in its FDN. Unlike \cite{nebula}, we do not estimate the storage or computational potential of our FDN, but the knowledge we maintain allows us to consider the cached video content of multiple FDN for every video being streamed.

Provensi \etal~\cite{maelstream} worked on a platform called Maelstream, a decentralized, self-organizing system that delivers media streams in a peer-to-peer manner. They focus on live streaming applications with users that consist of producers and consumers, such as webinars. Each node of Maelstream receives its media stream from neighboring nodes based on dynamic latency estimations and fair bandwidth usage. Similarly, F-FDN chooses the FDN from where videos are streamed based on accurate latency estimations in addition to the estimations from on demand video stream processing. Also, F-FDN follows a hybrid peer-to-peer and hierarchical structure as opposed to a purely decentralized nature that Maelstream has.

Many research works have been carried out to improve system performance utilizing fog computing (\eg~\cite{cloudfog,nebula}), however, none of them have concentrated on video streaming in the ways we propose in F-FDN.

\section{Conclusions and Future Work}\label{sec:conclsn}

We presented F-FDN, a novel distributed platform for low latency video streaming that enhances the traditional CDN system with fog computing capabilities. The primary goal of F-FDN is to deliver a streaming service with a high QoE, specifically to viewers located in geographically distant areas. Along with the proposal of the platform, we created a method to make intelligent streaming decisions at each FDN, so that the likelihood of having an uninterrupted video streaming experience is maximized. Our experiment results show F-FDN having an average of a 52\% improvement in deadline miss rate, when compared to the traditional CDN system that neither processes videos at the edge nor considers the cached video content of neighbors. Adaptive decision making within each FDN provides robustness against network latency fluctuations. To further enhance F-FDN, one future work will involve capturing other forms of uncertainty in network behavior. Another future work will be exploring a multi-tiered structure of FDNs within F-FDN and further sensitivity testing of our platform to measure the effect of additional parameters.

\section*{Acknowledgments}
We are thankful to anonymous reviewers of this paper. This research was supported by the Louisiana Board of Regents under grant number LEQSF(2016-19)-RD-A-25. 
%
\bibliographystyle{IEEEtran} 
 \bibliography{references}


\end{document}